# From Classical Rationality to Contextual Reasoning: Quantum Logic as a New Frontier for Human-Centric AI in Finance

Fabio Bagarello*    Francesco Gargano†    Polina Khrennikova*‡

October 7, 2025

**Abstract**

We consider state-of-the-art applications of artificial intelligence (AI) in modelling human financial expectations and explore the potential of quantum logic to drive future advancements in this field. This analysis highlights the application of machine learning techniques, including reinforcement learning and deep neural networks, in financial statement analysis, algorithmic trading, portfolio management, and robo-advisory services. We further discuss the emergence and progress of quantum machine learning (QML) and advocate for broader exploration of the advantages provided by quantum-inspired neural networks. These benefits arise from quantum logic's ability to capture agents' non-classical expectations and non-expected utility decisions, often referred to as *bounded rationality*. We present illustrative examples of expectation formation schemes in asset trading, grounded in quantum probability theory. We argue that quantum-based models hold significant potential to replicate human cognitive processes, enhance AI efficiency, and improve functionality in complex and uncertain environments. Ultimately, we aim to promote the adoption of quantum-driven AI techniques to improve upon classical models in capturing human-like decision-making.

**Keywords**: Artificial intelligence; Bounded rationality; Deep neural networks; Financial expectations; Investments; Quantum logic; Quantum machine learning

# Introduction

Artificial intelligence (AI) is a ubiquitous concept that frequently appears in media, policy papers, and university curricula. AI encompasses all fields where human intelligence can be substituted for automated processes. An algorithm performs some less (e.g., searching through a database) or more human-like tasks (executing trades against humans to maximize payoffs, analyzing complex and incomplete datasets to arrive at predictions, and making decisions). When discussing applications or problems that AI addresses in various scientific and economic domains, one notes specific algorithms that solve these problems as part of various machine learning techniques. In a nutshell, machine learning considers tasks of prediction that we can use in our decisions (which involve either the prediction of the expected values of continuous variables or the classification of variables with a finite number of realizations). An important role in contemporary economic and financial applications is played by the reinforcement learning (RL) technique, which involves the AI algorithm interacting with an outer environment. In such a dynamic context, RL processes feedback information to update policies that aim to maximize expected rewards, subject to constraints and the agent's anticipations about future states of financial markets. A central tool enabling this process are deep neural networks (DNNs), which build on earlier artificial neural network models but employ multiple hidden layers between input and output to more closely mimic the information processing of the human brain (see [58] for a detailed discussion).

*Dipartmento di Ingegneria, Universita di Palermo, 90128 Palermo, Italy and I.N.F.N., Sezione di Catania, Italy, email: fabio.bagarello@unipa.it

†Dipartmento di Ingegneria, Universita di Palermo, 90128 Palermo, Italy, email: francesco.gargano@unipa.it

‡*Corresponding author. Faculty of Behavioural, Management, and Social Sciences, Financial Engineering group, University of Twente, Netherlands, email: p.khrennikova@utwente.nl.

The conceptualization of computational neural networks can be traced back to the 1943 paper by [52] that paved the way to two avenues of research: *i)* the biological structures that would be found in a human brain and *ii)* the application of artificial neural networks as part of what is now known as contemporary AI. The paper by [33] revolutionized the wider application of neural network structures to deep learning (DL). It is now an established technique that can be used in complex learning environments where adaptations to constantly changing dynamic conditions are needed. One of the emerging domains in AI research is its application to finance, including areas such as algorithmic trading, financial advising, and risk management. These algorithms process historical market data, extract latent patterns, and generate predictions to guide trading strategies, financial advice, or portfolio management.

In this contribution, we stipulate a motivation from the perspective of quantum logic, for a wider exploration and implementation of quantum machine learning techniques, notably quantum neural networks, which are getting increased attention as more powerful tools to cope with complexity, non-linearity, unpredictability and high-dimensionality of learning environment, as discussed in the recent study by [1]. We draw upon the findings from behavioral economics, particularly the concept of bounded rationality, to elucidate the apparent deviations from rationality in human expectations and decision-making within complex and unfamiliar environments. Importantly, from the perspective of rational expectations and expected utility maximization in economics, such 'non-rational' behavior reflects a departure from the foundational axioms of classical probability theory and Bayesian reasoning, which underpin expected utility frameworks under conditions of both objective and subjective risk, [44, 73]. Against this backdrop, the so-called "biases" in human reasoning (that are often used as an umbrella for all non-expected utility-type behaviour) are gaining increasing attention as part of responsible and explainable AI, see, e.g., a survey of possible biases in applications to business decisions by [20]. While some cognitive biases can indeed be detrimental to decision-makers and other stakeholders, this analysis specifically focuses on biased information processing and expectation formation under conditions of incomplete information, rather than on biases viewed as inherently irrational or harmful. We aim to advocate for AI systems that more closely mirror human cognitive processes, particularly in terms of the underlying logic and neurological mechanisms. To achieve this, we draw on insights from the emerging field of 'quantum cognition,' which seeks to model human reasoning through the mathematical and logical frameworks of quantum physics, [16], [40], [60].

# Expectation Formation of Classical AI algorithms: Applications to Financial Decisions

One of the recent advancements in AI algorithms involves deep learning, particularly reinforcement learning, which seeks value-maximizing courses of action in dynamic financial market environments. These algorithms are especially well-suited to supporting investors in financial prediction and decision-making. Notably, they engage in a two-way exchange of information with the financial environment—such as prices, market and corporate news, and social media sentiment—closely mirroring the way human investors learn through trading, news analysis, and market evaluation. As a result, deep learning systems are increasingly applied both in advisory and active roles in portfolio management, including robo-advising, algorithmic trading, and customized chatbots. Financial institutions also leverage these systems to forecast and manage risks such as value at risk and probability of loss (for an introduction to AI algorithms in finance, see, e.g., the primer by [59]).

In active portfolio management, AI algorithms can potentially substitute for human decision-makers by aligning trades with individual preferences—such as maximizing the Sharpe ratio, minimizing drawdowns, or managing liquidity levels. It is important to note that active management strategies typically assume market inefficiencies, where excess returns can be captured from "hidden" or overlooked information. The evolution of AI in finance has been extensive, progressing from supportive tools to autonomous systems capable of sophisticated, cognitive decision-making, particularly in areas such as risk management and corporate strategy. These systems, including deep neural network-based algorithms, also engage in autonomous expectation-building by analyzing historical and real-time data to construct dynamic anticipatory models of risk and return [22].

Deep neural networks (DNNs) are trained on historical market data, which includes price movements deviating from the rational expectations equilibrium caused by investors' expectation-driven trading, [45]. Asset prices can exhibit unpredictable fluctuations, with stronger dependencies between financial asset prices than can be explained through classical correlation analysis, [25]. Furthermore, these AI algorithms may be trained in noisy, highly uncertain contexts where 'black swan' events are possible, such as unforeseen volatility shocks. Consequently, *predicting* expectation-driven trading behavior of investors becomes as essential as processing large datasets on historical prices and trading volumes, particularly in complex financial environments such as climate investments, cryptocurrencies, and derivatives.

When discussing deep learning and neural network applications in finance, the "input" data that yields final decisions—such as credit approval, fraud detection, portfolio construction, and risk management—is difficult to trace back, as the variety of data sources is immense and often incomplete or imbalanced across financial variables. In this regard, the primary ethical considerations regarding AI biases revolve around the inaccuracy of historical training data and the amplification of existing biases during the training of AI algorithms. Another concern is the learning process of DNNs, which involves numerous parameters that are difficult to trace back in the prediction process, resulting in a lack of explainability and the well-known "black box" issue. While biases are widely discussed in the literature on AI adaptation (e.g., discrimination against certain groups of borrowers in banking, skewed data sets, missing data, etc.), we would like to emphasize a slightly different manifestation of heuristics and biases, also coined "bounded rationality" in the seminal work of Simon [68]. In complex learning contexts, people seek to ease the cognitive load by reasoning with shortcuts and approximations [27]. Such cognitive mechanisms result in seemingly irrational human expectations and decisions from the perspective of classical logic [14].

As shown by behavioral finance scholars, the non-Bayesian update of information by investors can lead to suboptimal portfolio choices and subsequently a mispricing of financial assets. At the same time, investors value financial assets based on their future expectations, and hence, deviations from fundamental prices can persist for longer periods.[1]

It has been widely observed that investors tend to process macroeconomic and company-specific signals in a non-Bayesian fashion characterized by underreaction and a subsequent gradual and non-linear adjustment of future expectations about financial assets, as formalized by [7] and recently, for high-frequency trading by [35]. The literature on "sticky expectations" about macro and company-specific variables also considers a lack of belief adjustment as a form of underreaction to news, which can explain the predictability of stock prices. Experimental studies further demonstrate that deviations from Bayesian posterior updating can arise through different channels: the study by [48] documents asymmetric learning, where prior negative signals dampen responses to subsequent favorable information, resulting in underreaction; whereas [23] show that correlation neglect leads investors to treat redundant signals as independent, thereby overweighting them in belief formation and producing overreaction. These information-processing patterns contribute to systematic departures from rational expectations. For instance, stocks of companies with high profitability ratios consistently yield higher returns on a risk-adjusted basis, primarily because investors update profitability news incompletely, underweighting new information while extrapolating past profits and exhibiting a pessimistic bias for high-profit firms, [15].

The above cognitive mechanisms behind non-classical expectation formation are more pronounced in unfamiliar environments, showing that investors engage in a form of *adaptive learning* that is at variance with the normative appeal of information updates following the tenets of classical probability and rational expectations frameworks [14], [73]. Another issue with the Bayesian updating scheme is its linear structure, which prevents it from handling extreme events (such as probabilities of zero or one) and from capturing cases of evidence accumulation where the sequence of news affects the decision maker. For a detailed analysis, and for a quantum probability representation as a more general scheme of such information-processing cases, see, e.g., [8]-[9] and [26]. This discussion leads us to consider a different model of human (and AI) expectation formation based on quantum logic, which relaxes certain mathematical axioms of classical probability theory. In the next section, we will discuss the advances in quantum information in AI algorithms and briefly explore the distinguishing features of quantum logic [12].

---

[1]Following an important remark of the reviewer we would like to stress that the non-Bayesian update mode might be an evolutionary response to information processing under the complexity of environmental factors, and bounded rationality is a natural strategy to deal with uncertain and non-linear future, [36]-[37], and [68].

# Quantum Logic and Applications to AI

Quantum AI is closely associated with the advent of quantum computers. Richard Feynman advocated back in 1982 that quantum computation could be a superior solution to classical algorithms when dealing with high dimensionality and complexity. Henceforth, the field of quantum AI became closely associated with quantum computing and quantum algorithms (see [58] for a thorough treatment). The application of AI in fields such as finance, economics, language, and healthcare has grown tremendously over the past decade. However, this development occurred in parallel with advancements in quantum computing and machine learning, which were less focused on applications related to human prediction and decision-making. For a detailed review of historical developments and key advances in quantum algorithms, see, e.g., [77]. Quantum information studies were again recognized as ground-breaking with the award of the Nobel Prize 2022 for experimental and theoretical studies on the behaviour of quantum systems. The worldwide recognition of quantum phenomena, such as superposition and entanglement, in micro- and macroscopic systems closely followed renowned experimental work carried out by the Google corporation, demonstrating an absolute advantage of quantum computers in terms of a tremendous speedup of computations [2]. It was stated by the team that: *"The Google team estimates that simulating the full circuit would take 10,000 years even on a computer with one million processing units (equivalent to around 100,000 desktop computers). Sycamore [quantum computer name] took just 3 minutes and 20 seconds"*. p. 462.

Against this backdrop, the application of quantum calculus-based algorithms to information retrieval presents a promising approach to overcoming the limitations of classical AI algorithms rooted in symbolic logic, [53]-[54]. The evidence is now vast that quantum computation and quantum algorithms are capable of: i) speeding up computation and, ii) providing more accurate solutions to prediction and classification of outcomes in various domains ([13], [19], [51], [56]).[2] Some practical implementations of QML (quantum machine learning) and hybrid quantum-classical algorithms have also revolutionized their use in fields beyond physics. QML and quantum circuits were implemented to produce handwritten text [34], and to enhance the quality within ghost imaging [76]. The work by [18] furthered scalable quantum computing by developing a source that generates indistinguishable quantum photons for use in distributed quantum networks. The study by [10] advanced the architecture of quantum neural networks by utilizing quantum neurons as part of a feed-forward neural network structure. Recently, [50] developed a theoretical framework for quantum neural network parameter training to attain "over-parameterization," i.e., establishing a critical threshold of parameters at which adding more parameters prompts a leap in the neural network's performance during the training process.

Quantum machine learning (QML) emerges as an innovative application alongside classical AI algorithms and deep reinforcement learning approaches, [60]. One fundamental distinction of QML from classical supervised and unsupervised learning algorithms is its grounding in quantum, rather than classical, logic. The core differences lie in how bits in classical versus quantum computers are constructed; while a classical bit is *either* 1 or 0, a qubit can be a *superposition* of these two states. The interference of quantum states has the potential to significantly speed up algorithms, enabling them to process large volumes of information from dynamic environments, such as high-frequency financial data. Advancements in computational speed are evolving rapidly, with significant commercial breakthroughs, such as Microsoft's progress in topological superconductors and Google's innovations in quantum chips, laying the foundation for large-scale quantum computing, [24].

---

[2]The breakthroughs in quantum computing are not without controversy; see, e.g., [29] and [38], which point to statistical and methodological issues in Google's quantum supremacy experiment and call for further empirical and theoretical work to assess whether the reported speedup could be reproduced by classical methods under plausible assumptions. A detailed discussion of these assumptions lies beyond the scope of the present review, as our primary focus is on the parallels between the quantum logic underlying quantum algorithms and the properties of human cognition.

## Quantum Logic and its Properties

As discussed in section (1), there is a stream of research that aims to capture human reasoning and expectation formation with the aid of quantum probability calculus that obeys quantum logic, see among other analyses and reviews in [16]-[17], [40], [42], [46], [60], and [62]-[64]. The field emerged following the empirical findings of the behavioural economics school on the non-classicality of human behaviour, e.g., [37], [42], and [66].

Quantum probability relaxes several axioms of classical probability theory, including distributivity and commutativity, and consequently, the Bayesian formula for conditional probability. Moreover, quantum probability is grounded on Hilbert space representation, where events are represented by vectors in that space. Operationally, the relaxation of the above axioms enables the formalization of the coexistence of multiple expectations and choices simultaneously, when applied to human cognition. Put differently, the human brain can find itself in multiple (dynamically evolving) interwoven mental states [16]-[17], [42], [40]. In a recent paper by [41], it was demonstrated that the human cognitive system can exhibit quantum properties of indeterminism on a micro-level, as neurons can simultaneously occupy states of "firing" and "non-firing" associated with specific energy levels.

The phenomenon of indeterminism and coexistence is formally coined in quantum theory as *superposition* of states, and can well explain the 'boundedly rational' expectations about the future states of economy, expected dividends and prices of financial assets, as discussed in quantum probability frameworks of asset trading by e.g., [4]-[5], [32], [45]-[46]. The observed asset prices can deviate from the fundamental values based on perfect information and perfect foresight, as postulated by the rational expectations hypothesis (REH). The deviations from the uniquely possible equilibrium price states in financial markets emerge from human expectation-driven trading decisions. As shown in [46], the quantum probability-based models can explain the trading behaviours of investors, who exhibit heterogeneous beliefs and hence can 'agree to disagree on the future states of financial variables'. It can be debated whether non-classical reasoning is optimal for humans or should be corrected by AI algorithms, as seen in [3]. Here, we would like to discuss AI advancements that could support investors by mimicking their neural processes and yet outperforming them in terms of computational ability and speed. We consider human cognitive processes to be central to the subsequent advancement of 'intelligent' AI systems, as opposed to the normative rules of expectation formation based on a black-box approach; see also the analysis in [28].

## Classical and Quantum Probability Calculus

We outline examples of the classical and quantum logic structures that underpin the operation of AI algorithms, including the previously discussed deep neural networks and quantum neural networks. The main distinguishing property of classical logic (and classical probability calculus) is that probabilistic measurements obey the rules of commutativity and distributivity. Consider some events, $A$, $B$, and $C$, where $\{B, C\}$ form a disjoint partion of the sample space whence, $\{B \cup C\} = \Omega$, and $p(B) > 0$, $p(C) > 0$. We can express the probability ($p$) for an event $A$ as a union of its conjunctions with some arbitrary disjoint events $B$ and $C$. This gives the so-called "Formula of total probability".

$$p(A) = p(A \cap B) + p(A \cap C) \tag{1}$$

or in the form of conditional probabilities:

$$p(A) = p(A|B)p(B) + p(A|C)p(C) \tag{2}$$

In the quantum operational framework, this proposition may not hold, and this is precisely what occurs when a so-called "disjunction effect" in human decisions is observed [66]. The other notable deviations from classical logic give rise to order effects (violation of commutativity) and contextual preferences when the joint probability distribution cannot be observed. When the equality in Eq.(2) does not hold, one deals with non-additivity of probabilities (or non-additivity of the subjective expectations of an economic agent). The emergence of non-commutativity naturally involves a deviation from the formation of conditional expectations following the Bayesian update.

Experimental studies have shown that the non-distributivity and non-commutativity properties of human reasoning emerge when agents form and update beliefs in unfamiliar or complex environments. The conditions under which agents switch to non-classical information processing remain an intriguing and largely unexplored question, [16], with particular relevance in the context of financial markets, [46]. Analysis of both existing and new experimental data on quantum probability models consistently demonstrates the significant impact of unfamiliar settings on the "compartmentalization" of state space, which is facilitated through complementary variable-observables within the quantum framework [65].

The violation of classical probability axioms in human reasoning is of a contextual nature that closely aligns with the earlier notion of human bounded rationality. In the quantum probability framework, the non-distributive feature of human reasoning can be well captured by vector-based calculus, allowing for a superposition between expectations about events in question. The formula (2) is extended with an interference term (IT), which captures the non-additivity as mentioned above. Interference is a key feature of quantum probability, based on information processing with complex probability amplitudes rather than absolute probability values.

$$p(A) = p(A|B)P(B) + p(A|C)p(C) + IT \tag{3}$$

We note that the interference term arises from representing probabilities with complex-valued amplitudes, and it can exhibit both negative and positive (constructive or destructive) interference.

We see that, like human neural processes, a quantum logic-based neural network would function by employing the superposition states (characterised by interference effects) between the quantum bits, whereby instead of a path approach as employed in Turing-type algorithms grounded on symbolic logic and functionalism, a simultaneous propagation of information across multiple possible paths can be captured. Such a quantum-based neural network algorithm would form *non-deterministic* or *contextual* expectations that are compatible with human features such as heterogeneity and bounded rationality. Next, we provide some illustrative examples of contextual information processing in investment choices, followed by a discussion of the implications for quantum algorithms and an overview of some early results of their application. In particular, we highlight how quantum principles, such as complementarity and superposition, can be harnessed to capture the complexity and uncertainty of financial time series, as well as the volatile impact of upcoming news.

# Non-classical Expectations and Contextually in Financial Decisions: Some Illustrations

In this section, we aim to provide examples of non-classical expectation formation in investment settings that deviate from the tenets of classical logic but can be effectively captured using the projective calculus of quantum physics.

## Born rule of expectation formation under ambiguity

As discussed, investors often trade under imperfect information, finding themselves in ambiguous expectation states regarding the future price realisations of a risky asset. Let us denote the price observable $P$, for the asset $I$ represented by an operator having the vectors $|\pm\rangle$ as eigenvectors with eigenvalues $\pm 1$ :

$$P = |+\rangle\langle+| \ - \ |-\rangle\langle-| \equiv P_+ - P_-. \tag{4}$$

Here, $|+\rangle$ and $|-\rangle$ respectively describe situations in which the price increases or decreases. To resolve the (non-classical) ambiguity, the agent measures the corresponding price expectation observable. Before forming a concrete preference regarding holding or selling the unit of an asset, the agent must resolve her uncertainty about the possible behavior of the stock price in the next trading period. This resolution of uncertainty can occur when new information about fundamentals or macroeconomic indicators becomes available. The Born rule gives the probability of the realization of a representative agent's ambiguity:

$$p(\pm) = |\langle\pm|\psi\rangle|^2 = \|P_\pm \psi\|^2. \tag{5}$$

Born's rule specifies the probability of obtaining a particular measurement result on the initial state. For simplicity, in this example, we only consider a pure state $\psi$ as the initial belief state of a representative investor. The formulation of the Born rule of state update differs for pure and mixed states; however, in both cases, it specifies the classical probability distribution associated with the price realisation of the $P$ observable. As such, the Born rule normalises the quantum measurement scheme with respect to an observable.

Such an expectation formation framework is in accord with the "bounded rational" account of investors' heuristic trading decisions under deeper uncertainty regimes in financial markets, when classical probabilities cannot be estimated for future price outcomes.

The rise of opaque asset classes, characterised by limited or no history of past returns, such as cryptocurrencies, contributes to the complexity of risks faced by today's investors. In particular, if the time evolution of a stock market is described, as it is natural in this quantum-like description, in terms of some Hamiltonian operator $H$, formula (5) can be easily extended to multi-period expectation formation:

$$p(\pm)[t] = |\langle\pm|\psi(t)\rangle|^2 = |\langle\pm|e^{-iHt}\psi\rangle|^2. \tag{6}$$

We observe that $\psi(t)$ remains a linear combination of the $|\pm\rangle$ vectors, with time-dependent coefficients that describe the influence of $H$ on the asset's price. Consequently, this can induce variations in the agent's measurement effect. It is important to emphasize that $H$ may also encompass the impact of various sources of information, which contribute to the dynamic changes in the asset's price or other market aspects [4].

## State Dependence in Asset Trading and Feedback Reaction

It has been shown that investors exhibit narrow framing concerning the evaluation of risky asset returns by treating future investment periods as complementary in the process of expectation formation [30], [49], [75]. Quantum probability-based measurement provides a natural way to capture such state dependence in expectations, where joint evaluation of prices and returns is absent. Building on this, and consistent with experimental studies on asset return feedback under various investment horizons, we consider discrete trading periods (illustrated with the case of two periods).

When anticipating asset price $\alpha_t = \pm 1$ at time $t$, the investor's state $\psi$ is projected onto the eigenvector $|\alpha_t\rangle$ that corresponds to an eigenstate for a particular price realization for that asset at that trading period, $t$. As the price at time $t$ is observed (e.g., price went up), the investor forms expectations about the possible price distribution of the asset at time $t+1$, and performs a measurement of the corresponding expectation observable, for the updated state $|+_{t+1}\rangle$. The eigenvalues of the price behaviour observable $P_t$, $\alpha_t = \pm 1$ are given via the state transition probabilities:

$$p_{t \to (t+1)}(\alpha_t \to \alpha_{t+1}) = |\langle\alpha_t|\alpha_{t+1}\rangle|^2. \tag{7}$$

The observed price outcome at time $t$ alters expectations about the asset price distribution at $t+1$, indicating state dependence in expectations, such as a belief in price trends. In general:

$$p_{t,(t+1)}(\alpha_t, \alpha_{t+1}) \neq p_{(t+1),t}(\alpha_{t+1}, \alpha_t). \tag{8}$$

The complementarity of expectations regarding the price observations of risky assets, $i_1...i_n$, can also manifest in portfolio selection and diversification decisions. We illustrate this with an example involving two risky assets, $i$ and $j$. An investor uncertain about the price dynamics of these assets lacks a *joint probability evaluation* of their price realizations. Interference effects arise concerning their potential price dynamics. Due to the incompatibility of the price observables, the investor considers the two assets sequentially, leading to order effects in expectation formation. This phenomenon is well-documented in the literature on order effects in judgments using quantum probability, e.g., [16], [43], and [74]. In an investment setting, the experimental study by [11] examines the effect of aggregated versus disaggregated portfolio performance reporting on investors' risk-taking in investment preferences. The study contributes to the extensive body of behavioural finance literature on feedback delays and isolated feedback on return expectations, with implications for investment preferences. We formalise the sequential information processing of assets that an investor trades as follows. By forming an expectation concerning the price up or down, $\alpha = \pm 1$ for the asset $i$, an investor's state $\psi$ is projected onto the eigenvector $|\alpha_i\rangle$ that corresponds to an eigenstate for a particular price trend. In the simple setup with two types of discrete price movements, there are only two eigenvectors $|\alpha_+\rangle$ and $|\alpha_-\rangle$, with eigenvalues $a = \pm 1$ corresponding to an upward or downward price movement of the financial asset. In the next trading period, the investor proceeds by forming an expectation about the possible price behaviour of the asset $j$. The investor starts her evaluation from an updated state of expectations, $|+_i\rangle$. Subsequently, her state transition concerning the price behaviour of the asset $j$ with eigenvalues $\beta = \pm 1$ is given by the Born rule, with the transition probabilities:

$$p_{i \to j}(\alpha \to \beta) = |\langle \alpha_i | \beta_j \rangle|^2. \tag{9}$$

The eigenvalues correspond to the tendency of the price to increase or decrease for the considered risky assets. In this framework, expectation formation about future asset prices occurs under ambiguity that is resolved sequentially. When forming an expectation about the first asset, the investor is in a superposition regarding the possible price behavior of the complementary asset, and interference effects emerge. Figure 1 presents a geometric representation of the sequential expectation formation scheme, illustrated through the example of two traded assets.

The above scheme of state transition yields the *quantum transition probabilities* that denote a representative investor's expectations concerning the asset $j$'s price distribution, after first observing the price behavior of asset $i$. These transition probabilities also have an objective, frequency interpretation. Consider an ensemble of investors in the same initial state of expectations, $\psi$, who make a decision $\alpha$ regarding the price behavior of the first asset, $i$. Subsequently, the agents form expectations about the $j$th asset. By considering only those investors whose expectation is $\beta$, it is possible to determine a frequency-probability $p_{i \to j}(\alpha \to \beta)$. Such quantum probabilities can be interpreted as analogues of classical conditional probabilities, $p_{i \to j}(\alpha \to \beta) \equiv p_{j|i}(\beta|\alpha)$.

Given these probabilities, we can define a quantum joint probability distribution for forming expectations about both assets $i$ and $j$:

$$p_{ij}(\alpha, \beta) = p_i(\alpha) p_{j|i}(\beta|\alpha). \tag{10}$$

For such a joint probability representation, the order structure is pivotal:

$$p_{ij}(\alpha, \beta) \neq p_{ji}(\beta, \alpha). \tag{11}$$

Sequential information processing manifests as *order effects* or state dependence in expectation formation, which does not align with classical Bayesian probability updates (see, e.g., [64], [74]). Order effects correspond to the non-satisfaction of the joint probability distribution and a violation of the commutativity principle central to classical probability theory [47]. This non-commutativity effect has an interesting consequence: a quantum-like approach intrinsically includes the possibility of considering any kind of *decision under uncertainty* as an effect of the Heisenberg-Robinson inequality for non-commuting operators [6]. Uncertainties naturally arise in this non-classical approach, eliminating the need to introduce them manually.

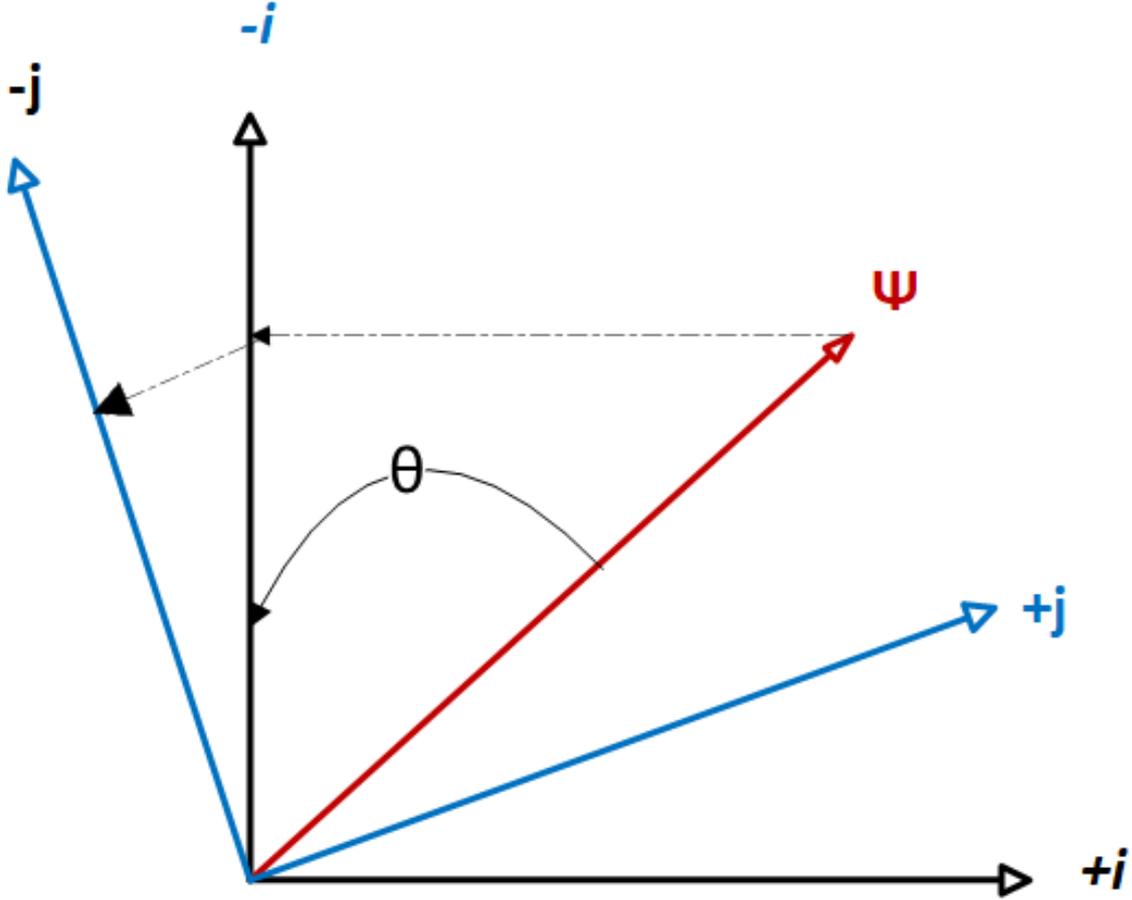

Figure 1: Geometric representation of an investor's state transition: starting from an initial trading state $\psi$, the investor sequentially evaluates asset price dynamics of the traded assets as complementary observables. For example, upon observing a price drop in asset $i$ (associated with eigenvalue $-i$), she updates her belief state such that a price drop in asset $j$ becomes more probable than it was under the prior state $\psi$.

The distinguishing feature of quantum probability models in financial expectation formation is the absence of a joint distribution of expectations regarding the future price behavior of traded assets (under certain conditions, as discussed earlier). This has significant implications for stock valuations and deviations from the rational expectations equilibrium.

# Quantum AI in financial models: some first results

Building on the conceptual foundations outlined above, we examine how emerging quantum AI techniques—including quantum neural networks (QNNs), quantum-enhanced reinforcement learning, and quantum machine-learning primitives such as quantum amplitude estimation (QAE) and its shallow-circuit variants—can be deployed in financial modeling to better capture non-classical investor expectations and accelerate the computation of pricing and risk metrics. Shallow-circuit variants can be implemented via variational quantum circuits, which are particularly suitable for today's computational hardware and enable preference- and expectation-consistent scenario analysis and stress testing under deep uncertainty and high-volatility regimes. Recent work by [70]-[71], following the studies by [63], shows quadratic error-scaling advantages for market-risk computation and gradients, and practical pathways via variational implementations suitable for today's computational hardware[3]. Hybrid QNNs based on variational quantum circuits have been implemented by [72] for forecasting stock index prices within the Chinese stock market, with evidence of performance gains over classical baselines. Such models could be trained to learn investor-specific contextual preferences (e.g., loss aversion, ESG tilts) and to detect regime shifts where classical stationarity assumptions break, thereby better matching stakeholder expectations in volatile and uncertain trading environments. More specifically, a Hilbert state space representation of financial time series (via phase state reconstruction) allows for capturing complex correlations and superpositions to deal with hidden and non-linear dependencies in financial time series. Quantum-enhanced reinforcement learning (QRL) offers a principled approach to encoding investors' risk-sensitive preferences and learning sequential allocation strategies that adapt to evolving market conditions and partial observability. Recent work in portfolio optimization by [69] integrates QNNs with evolutionary meta-reinforcement learning, demonstrating how QRL architectures can enhance context-awareness and adaptability across market regimes. Their framework captures non-stationary dynamics and investor-specific objectives, enabling more responsive and preference-aligned policy learning in complex trading environments. We can witness that the interest to financial applications of quantum algorithms and quantum computing is growing, with promising areas in risk management, quantum amplitude estimation for Monte-Carlo simulation, broader application of quantum computing in training of neural networks, quantum reinforcement learning, etc., (see also analyses in [13], [21] and [61]). The study by [55] demonstrates that utilising neuroevolution to train both the configurations and structures of neural networks, combined with the linearization of financial data through Principal Component Analysis (PCA), can also significantly enhance trading performance in the dynamic environment of financial markets.

These initial applications suggest that quantum-inspired approaches are not only conceptually valuable, but may also offer practical pathways for embedding investor preferences into financial decision-making.[4]

## Discussion and Conclusion

In this state-of-the-art paper, we aimed to identify the necessary conditions for employing AI systems that utilise quantum-inspired information processing to naturally capture the human neural processes that occur in the financial marketplace. We informed our discussion by comparing normative frameworks for preference formation based on unlimited cognitive ability and time, as postulated under REH, and bounded rationality, which relaxes the assumptions on how, e.g., investors form their expectations with limited existing knowledge and complex information. The use of quantum logic is well motivated and has been shown to better represent the nature of human reasoning in realistic decision-making (widely known under the umbrella of "Quantum Cognition" studies). Human neural processes can exhibit quantum features, such as the coexistence and strong correlations between these neural states. This is why quantum AI could be a natural advancement of classical AI algorithms to deliberate and form predictions in complex and uncertain environments, serving as a helpful aid for humans in various domains of financial decision-making, see also a report by [79] outlining the recent developments in financial applications. Against this backdrop, the core argument of this analysis is that AI could become a valuable human-like apparatus that allows us to capture the boundedly rational mechanism of human reasoning and equips us with better tools to deal with the high complexity and multidimensionality of financial data. Building on the initial results in quantum AI, a promising avenue for future research lies in integrating evolutionary algorithms with quantum logic-based deep neural networks (DNNs), thereby advancing the development of artificial general intelligence (AGI) systems that can more accurately reflect human expectations and investment preferences. Incorporating quantum features such as entanglement and superposition within these architectures may further enhance the explainability of AI models for boundedly rational investors, fostering closer correspondence with their expectation formation and investment behavior.

---

[3]Variational quantum circuits with gradient-based optimization can be considered as conceptually close to QNNs.

[4]For a state-of-the-art overview of quantum algorithm applications in investment settings, including portfolio optimization, risk assessment, and blockchain development, see [57].

# Acknowledgements

We gratefully acknowledge the valuable comments and insights from participants of the "Quantum Information and Probability: From Foundations to Engineering" (QIP24) conference, which greatly contributed to improving this paper. We also extend our special thanks to the participants, and especially to the organizer, Emmanuel Haven, of the special session on *Quantum Methods in Economics and Finance* for their thoughtful feedback and engaging discussion. P.K. acknowledges the support of "Digital Finance - Reaching New Frontiers" (Horizon Marie Sklodowska-Curie Actions Industrial Doctoral Network), Horizon Europe research and innovation, No. 101119635. F.B. and F.G. acknowledge support under the PNRR project funded by the European Union - NextGenerationEU - Project Title "Transport phonema in low-dimensional structures: models, simulations and theoretical aspects"- project code 2022TMW2PY - CUP B53D23009500006. F.B. and F.G. also acknowledge support from the FFR2024-FFR2025 grant of the University of Palermo and support of the G.N.F.M. of the INdAM. F.G. and P.K. acknowledge partial financial support under the "Networking" project of the Department of Engineering of the University of Palermo. F. B. was also supported by project ICON-Q, Partenariato Esteso NQSTI - PE00000023, Spoke 2. F.G. also acknowledges support from the PNRR Project QUANTIP – Partenariato Esteso NQSTI, PE00000023, Spoke 9, CUP: E63C22002180006.